# Finite Volume Model to Study the Effect of ER flux on Cytosolic Calcium Distribution in Astrocytes

Brajesh Kumar Jha, Neeru Adlakha and M. N. Mehta

**Abstract**— Most of the intra-cellular events involved in the initiation and propagation phases of this process has now been identified astrocytes. The control of the spread of intracellular calcium signaling has been demonstrated to occur at several levels including IP3 receptors, intracellular $Ca^{2+}$ stores like endoplasmic reticulum (ER) . In normal and pathological situations that affect one or several of these steps can be predicted to influence on astrocytic calcium waves. In view of above a mathematical model is developed to study interdependence of all the important parameters like diffusion coefficient and influx over $[Ca^{2+}]$ profile. Model incorporates the ER fluxes like $J_{leak}$, $J_{Pump}$ and $J_{Chan}$. Finite volume method is employed to solve the problem. A program has been developed using in MATLAB 7.5 for the entire problem and simulated on an AMD-Turion 32-bite machine to compute the numerical results. In view of above a mathematical model is developed to study calcium transport between cytosol and ER.

—————————— ◆ ——————————

## 1 INTRODUCTION

LARGE and long-lasting cytosolic calcium signalling in astrocytes have been described in cultured cells and acute slice preparations. The mechanisms that give rise to these calcium events have been extensively studied in vivo. However, their existence and functions in the intact brain are unknown [1],[6],[7][11]. About 20 years ago astrocytes have been considered to mediate supportive and protective functions in the central nervous system because of their strategic placement relative to the vasculature, and because they lack fast sodium action potentials. Now Astrocytes are controlling the dynamics of the neuronal networks in the central nervous system[8],[10].

In cultured astrocytes response to many physiological and pharmacological manipulations, for example- mechanical stimulation, membrane potential depolarization, and activation of metabotropic glutamate receptors etc. [2],[5],[11].

These slow events are mediated by release of $Ca^{2+}$ from intracellular stores (Charles et al. 1993)[4].The ER is the major $Ca^{2+}$ storage organelle in most cells. ER membrane $Ca^{2+}$ ATPases accumulate $Ca^{2+}$ in the ER lumen to quite high levels. Because the ER lumen contains high concentrations of $Ca^{2+}$ binding proteins, due to this total amount of $Ca^{2+}$ in the lumen may be >1 mM[7]. The concentration of $Ca^{2+}$ in the cytoplasm of unstimulated cells is 3–4 orders of magnitude lower than in the ER lumen. This low concentration is maintained by $Ca^{2+}$ pumps and other $Ca^{2+}$ transporters located in the ER, as well as plasma, membranes. Due to binding of InsP3, it provides a pathway for $Ca^{2+}$ to diffuse from the ER lumen to cytoplasm. $Ca^{2+}$ in the cytoplasm moves by passive diffusion. The $Ca^{2+}$ concentration adjacent to the open channel may be 100 μM or more, whereas concentrations as close as 1–2 μm from the channel pore may be below 1 μM [7],[9],[12]. Thus $Ca^{2+}$ signalling appears to provide the most versatile and crucial mechanisms for such an Astrocytes contribution.

Astrocytes posses a specific and significant route for $Ca^{2+}$ entry which is controlled by the filling state of the ER calcium stores. The De Young-Keizer model is the most general model that incorporates a scheme for the gating kinetics of the channels. The effect of calcium binding to the IP3-receptors was ignored on the calcium dynamics. Othmer-Tang et. Al. Simplify the De Young-Keizer model which exhibits both excitability and frequency encoding.

In view of above a mathematical model is developed to study cytosolic calcium profile for Astrocytes. To derive the calcium dynamics equation we follow the De Young-Keizer model. In proposed model the only calcium exchange taken into account is the one across the endoplasmic reticulum (ER). The model has been developed for a one dimension steady state case. The finite volume method [2],[8] is employed to obtain the solution. Calcium sequestration by the buffering proteins like calmodulin and exchanges across the mitochondria are not considered. Also, there is no calcium flux between the cell and the extra cellular medium, which implies that the total calcium concentration inside the cell is constant [5],[13],[14].

## 2 Mathematical Formulation

The conservation equation for calcium profile [14]

$$\eta_{Ca} = \eta_{ER} + \eta_{Cyt} \tag{1}$$

———————————————

- *Brajesh Kumar Jha is with the S. V. National Institute of Technology, Surat, India-395007*
- *Neeru Adlakha. is with the S. V. National Institute of Technology, Surat, India-395007*
- *M.N.Mehta is with the S. V. National Institute of Technology, Surat, India-395007*

Where $\eta_{Ca}$, $\eta_{ER}$ and $\eta_{Cyt}$ represent the number of moles of calcium inside the cell, the ER and the cytosol. Dividing both sides by the cytosolic volume $V_{Cyt}$ we get

$$\frac{\eta_{Ca}}{V_{Cyt}} = \frac{\eta_{ER}}{V_{ER}} \frac{V_{ER}}{V_{Cyt}} + \frac{\eta_{Cyt}}{V_{Cyt}} \qquad (2)$$

The conservation equation is given as

$$C_o = C_1 [Ca^{2+}]_{ER} + [Ca^{2+}]_i \qquad (3)$$

$$C_0 = \frac{\eta_{Ca}}{V_{Cyt}}, \quad C_1 = \frac{V_{ER}}{V_{Cyt}}, \quad [Ca^{2+}]_{ER} = \frac{\eta_{ER}}{V_{ER}}$$

and $\quad [Ca^{2+}]_i = \frac{\eta_{Cyt}}{V_{Cyt}} \qquad (4)$

Where $C_0$ is 'total' calcium concentration (with respect to the cytosolic volume), $C_1$ is the volume ratio between the ER and the cytosol and $[Ca^{2+}]_i$ is the cytosolic concentration. The three fluxes are incorporated for all the calcium fluxes between the ER and cytosol are $J_{leak}$, $J_{Chan}$ and $J_{Pump}$.

$$\frac{dn_i}{dt} = J_{leak}.V_{ER} + J_{Chan}.V_{ER} - J_{Pump}.V_{Cyt} \qquad (5)$$

Where the different fluxes are expressed in $\mu M.\sec^{-1}$. The material balance equation in terms of concentration using notation (4) is written as

$$\frac{d[Ca^{2+}]_i}{dt} = C_1 [J_{leak} + J_{chan}] - J_{Pump} \qquad (6)$$

This is the general form of the calcium dynamics equation. Here it is also assumed that the calcium fluxes occur solely between the ER and the cytosol. The outward fluxes ($J_{leak}$ and $J_{chan}$) are obtained from Fick's law of diffusion.

$$J_{leak} = D_{leak} \left( [Ca^{2+}]_{ER} - [Ca^{2+}]_i \right)$$

$$= D_{leak} \left( \frac{C_0 - [Ca^{2+}]_i}{C_1} - [Ca^{2+}]_i \right) \qquad (7)$$

$$= \frac{D_{leak}}{C_1} (1 + C_1) \left( \frac{C_0}{1 + C_1} - [Ca^{2+}]_i \right)$$

$$J_{Chan} = P.\frac{Channels}{unitvolER * unittime}.V_{ER}.\left( [Ca^{2+}]_{ER} - [Ca^{2+}]_i \right) \qquad (8)$$

or equivalently,

$$J_{Chan} = P.D_{Chan}.\left( [Ca^{2+}]_{ER} - [Ca^{2+}]_i \right)$$

$$= \frac{D_{Chan}}{C_1} (1 + C_1) \left( \frac{C_0}{1 + C_1} - [Ca^{2+}]_i \right) \qquad (9)$$

Where $D_{leak}$ is the constant of diffusivity associated to the leak and $D_{Chan}$ is the channel conductance expressed in sec$^{-1}$. P is the open channel probability (proportional to the number of channels open).

The inward flux $J_{Pump}$ is modelled by a Michaelis-Menten expression with a Hill coefficient 2

$$J_{Pump} = P_R^{\max} \frac{[Ca^{2+}]_i^2}{[Ca^{2+}]_i^2 + (K_R^M)^2} \qquad (10)$$

Where $P_{R_1}^{\max}$ is the maximal pump rate expressed in $\mu M.\sec^{-1}$, and $K_R^M$ is the Michaelis-Menten constant in $\mu M$. Finally the general form of the calcium dynamics equation using Fickian diffusion can be stated as

$$\frac{\partial [Ca^{2+}]_i}{\partial t} = D_{Ca} \frac{\partial^2 [Ca^{2+}]_i}{\partial x^2} + (1+C_1)(D_{leak} + P.D_{Chan}).\left( \frac{C_0}{1+C_1} - [Ca^{2+}]_i \right)$$

$$- P_R^{\max} \frac{[Ca^{2+}]_i^2}{[Ca^{2+}]_i^2 + (K_R^M)^2} \qquad (11)$$

for a steady state case the equation (8) is reduced in the form :

$$D_{Ca} \frac{d^2 C}{dx^2} + (1+C_1)(D_{leak} + P.D_{Chan}).\left( \frac{C_0}{1+C_1} - C \right)$$

$$- P_R^{\max} \frac{C^2}{C^2 + K^2} = 0 \qquad (12)$$

Along with boundary conditions as:

$$\lim_{x \to 0} \left( -D_{Ca} \frac{d[Ca^{2+}]}{dx} \right) = \sigma_{Ca} \qquad (13)$$

$$\lim_{x \to \infty} [Ca^{2+}] = 0.1 \mu M \qquad (14)$$

To handle the nonlinear term two cases are considered

**CASE-I**

When $K \gg C$

Then $\quad \dfrac{C^2}{C^2 + K^2} < \dfrac{C^2}{K^2} < \dfrac{C}{K}$

Thus equation (12) reduced as

$$D_{Ca}\frac{d^2C}{dx^2}+(1+C_1)(D_{leak}+P.D_{Chan})\cdot\left(\frac{C_0}{1+C_1}-C\right)$$
$$-P_R^{max}\frac{C}{K}=0 \quad (15)$$

In order to apply the finite volume method the domain is divided into discrete control volumes (Figure 1).

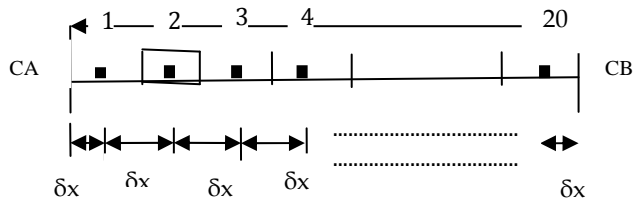

Fig. 1 Domain Discretization

Taking 20 nodal points in the space between A and B. Each node is surrounded by a control volume or cell. A general nodal point is identified by P and its neighbours in a one-dimensional geometry, the nodes to the west and east, are identified by W and E respectively. The west sides face of the control volume is referred by w and the east side control volume face by e. The distances between the nodes W and P, and between nodes P and E, are identified by $\delta x$. Similarly the distance between face w and point P and between P and face e are denoted by $\delta x/2$. Nodal values to the east and west are available at nodal values 2, 3, 4............19. For the simplicity equation (15) can be written as

$$\frac{d^2C}{dx^2}-p_1C+q_1=0 \quad (16)$$

Where

$$p_1=\frac{1}{D_{Ca}}\left((1+C_1)(D_{leak}+P.D_{Chan})+\frac{P_R^{max}}{K}\right)$$
$$q_1=\frac{C_0}{D_{Ca}}(D_{leak}+P.D_{Chan}) \quad (17)$$

The calcium concentration $[Ca^{2+}]$ is replaced by C for convenience. Integration of equation (16) over control volume gives [2],[8]:

$$\int_{\Delta V}\frac{d^2C}{dx^2}dV-p_1\int_{\Delta V}CdV+q_1\int_{\Delta V}dV=0 \quad (18)$$

$$\left[\left(A\frac{dC}{dx}\right)_e-\left(A\frac{dC}{dx}\right)_w\right]-p_1C_pA\delta x+q_1A\delta x=0 \quad (19)$$

$$\left(\frac{dC}{dx}\right)_e-\left(\frac{dC}{dx}\right)_w-p_1C_p\delta x+q_1\delta x=0 \quad (20)$$

$$\left(\frac{C_E-C_P}{\delta x}\right)-\left(\frac{C_P-C_W}{\delta x}\right)-p_1C_p\delta x+q_1\delta x=0 \quad (21)$$

This can be rearranged as

$$\left[\frac{1}{\delta x}+\frac{1}{\delta x}+p_1\delta x\right]C_P=\frac{1}{\delta x}C_W+\frac{1}{\delta x}C_E+q_1\delta x \quad (22)$$

The general form for the interior nodal point 2, 3, 4........19 is given by

$$a_PC_P=a_WC_W+a_EC_E-S_u \quad (23)$$

$$a_W=\frac{1}{\delta x},\ a_E=\frac{1}{\delta x},a_P=a_W+a_E-S_P, \quad (24)$$
$$S_P=-p_1\delta x,\ and\ S_u=q_1\delta x$$

We apply the boundary conditions at node points 1 and 20. At node 1 west control volume boundary is kept at specified concentration

$$a_P=a_W+a_E-S_P,\ a_W=0,\ a_E=\frac{1}{\delta x},$$
$$S_P=-\left(\frac{2}{\delta x}+p_1\delta x\right),\ and\ S_u=\left(\frac{2}{\delta x}\right)C_A+q_1\delta x \quad (25)$$

Similarly at node 20 east control volume boundary is at specified concentration.

$$a_P=a_W+a_E-S_P,\ a_W=\frac{1}{\delta x},\ a_E=0,$$
$$S_P=-\left(\frac{2}{\delta x}+p_1\delta x\right),\ and\ S_u=\left(\frac{2}{\delta x}\right)C_B+q_1\delta x \quad (26)$$

In equation (23), putting all values of equation (24-26) we have a system of algebraic equations as given below. Where $C_A$ and $C_B$ be the specified boundary conditions in terms of calcium concentration.

$$AX=B \quad (27)$$

Here, $X=c_1,c_2................c_{20}$ represents the calcium concentration, A is system matrices and B is the system vector.

**CASE-II**

When $K<<C$
Putting $K=\alpha C$ for $0<\alpha<1$

Then $\frac{C^2}{C^2+K^2}=\frac{1}{\alpha^2+1}$ for $0<\alpha<1$

Thus equation (12) reduced as

$$D_{Ca}\frac{d^2C}{dx^2}+(1+C_1)(D_{leak}+P.D_{Chan})\cdot\left(\frac{C_0}{1+C_1}-C\right)$$
$$-\frac{P_R^{max}}{\alpha^2+1}=0 \quad (28)$$

For the simplicity equation (28) can be written as

$$\frac{d^2C}{dx^2}-p_2C+q_2=0 \quad (29)$$

Where

$$p_2=\frac{1}{D_{Ca}}(1+C_1)(D_{leak}+P.D_{Chan})$$

$$q_2=\frac{1}{D_{Ca}}\left(C_0(D_{leak}+P.D_{Chan})-\frac{P_R^{max}}{\alpha^2+1}\right) \quad (30)$$

The finite volume scheme is employed to solve equations (29) together with (13)-(14).

$$\int_{\Delta V}\frac{d^2C}{dx^2}dV-p_2\int_{\Delta V}CdV+q_2\int_{\Delta V}dV=0 \quad (31)$$

$$\left[\left(A\frac{dC}{dx}\right)_e-\left(A\frac{dC}{dx}\right)_w\right]-p_2C_pA\delta x+q_2A\delta x=0 \quad (32)$$

$$\left(\frac{dC}{dx}\right)_e-\left(\frac{dC}{dx}\right)_w-p_2C_p\delta x+q_2\delta x=0 \quad (33)$$

$$\left(\frac{C_E-C_P}{\delta x}\right)-\left(\frac{C_P-C_W}{\delta x}\right)-p_2C_p\delta x+q_2\delta x=0 \quad (34)$$

This can be rearranged as

$$\left[\frac{1}{\delta x}+\frac{1}{\delta x}+p_2\delta x\right]C_P=\frac{1}{\delta x}C_W+\frac{1}{\delta x}C_E+q_2\delta x \quad (35)$$

The general form for the interior nodal point 2, 3, 4........19 is given by

$$a_PC_P=a_WC_W+a_EC_E-S_u \quad (36)$$

$$a_W=\frac{1}{\delta x},\ a_E=\frac{1}{\delta x},a_P=a_W+a_E-S_P,$$
$$S_P=-p_2\delta x,\ and\ S_u=q_2\delta x \quad (37)$$

We apply the boundary conditions at node points 1 and 20. At node 1 west control volume boundary is kept at specified concentration

$$a_P=a_W+a_E-S_P,\ a_W=0,\ a_E=\frac{1}{\delta x},$$
$$S_P=-\left(\frac{2}{\delta x}+p_2\delta x\right),\ and\ S_u=\left(\frac{2}{\delta x}\right)C_A+q_2\delta x \quad (38)$$

Similarly at node 20 east control volume boundary is at specified concentration.

$$a_P=a_W+a_E-S_P,\ a_W=\frac{1}{\delta x},\ a_E=0,$$
$$S_P=-\left(\frac{2}{\delta x}+p_2\delta x\right),\ and\ S_u=\left(\frac{2}{\delta x}\right)C_B+q_2\delta x \quad (39)$$

In equation (29), putting all values of equation (36-39) we have a system of algebraic equations as given below. Where $C_A$ and $C_B$ be the specified boundary conditions in terms of calcium concentration.

$$AX=B \quad (40)$$

Here, $X=c_1,c_2\ldots\ldots\ldots\ldots c_{20}$ represents the calcium concentration, A is system matrices and B is the system vector.

## 3. Results and Discussion:

The numerical results for calcium profile against different biophysical parameters have been obtained using numerical values of parameter given in table 1 unless stated along with figures.

**Table I**
**List of physiological parameters used for numerical results**

| Symbol | Parameters | Values |
|---|---|---|
| $C_1$ |  | 0.185 |
| $P_R^{max}$ | Maximum $Ca^{2+}$ uptake ($\mu M.sec^{-1}$) | 0.9 |
| $D_{leak}$ | $Ca^{2+}$ leak flux constant (sec$^{-1}$) | 0.11 |
| $D_{Chan}$ | Channel conductance (sec$^{-1}$) | 6 |
| $K_R^M$ | Dissociation constant of $Ca^{2+}$ to the pump ($\mu M$) | 0.1 |
| D | Average $Ca^{2+}$ concentration ($\mu M$) | 1.688 |

The numerical results for calcium profile against different biophysical parameters have been obtained using numerical values of parameter given in table 1 unless stated along with figures.

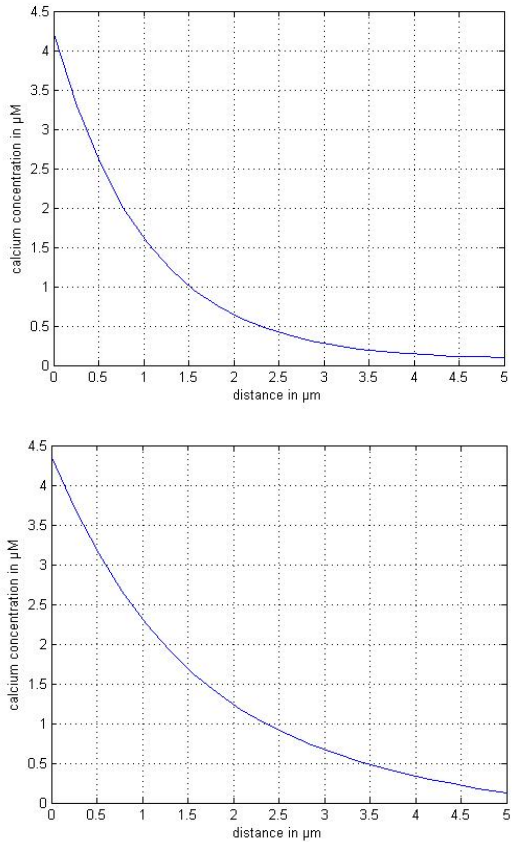

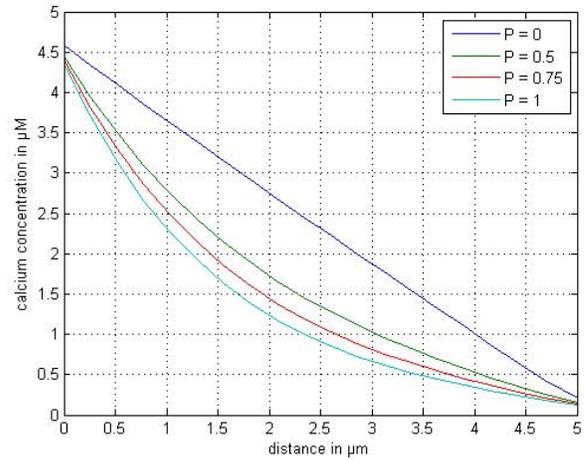

Figure 3 (a) and 3 (b) spatial variation of calcium concentration number of channel varies closed the calcium concentration remain little higher than the channel become open. Due to different number of channel open calcium profile become low as more number of channel become opened.

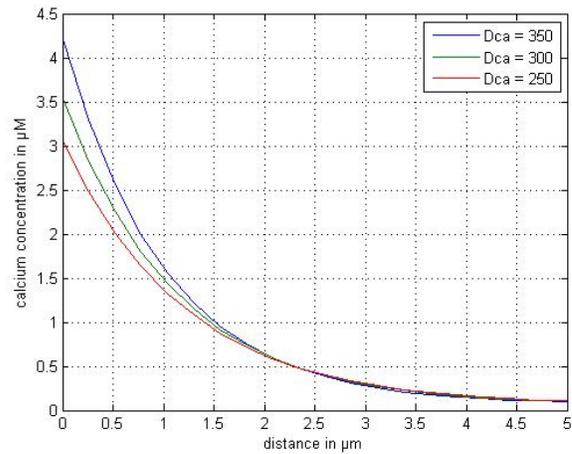

Figure 2 (a) and 2 (b) spatial variation of calcium concentration

Figure 2 (a) and 2 (b) shows the spatial variation of calcium In Case I calcium concentration start from 4.5 μM and goes down rapidly up 0.5μM at 3.5μm. In Case I calcium concentration decrease more rapidly than case II then after become constant as we move far from the source.

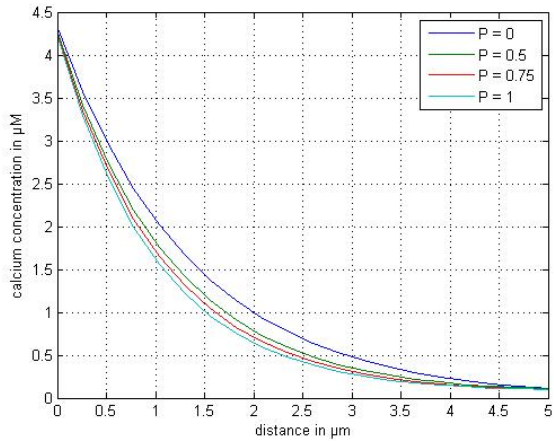

Figure 3 (a) and 3 (b) shows the spatial variation of calcium concentration when number of calcium channel between cytosol and ER. Since the calcium concentration is more high than in cytosol so as calcium channel remain

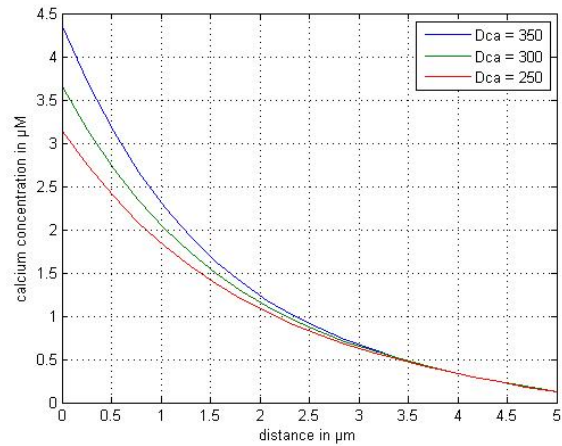

Figure 4 (a) and 4 (b) spatial variation of calcium concentration for different value of Diffusion coefficient

Figure 4 (a) and 4 (b) shows the spatial variation of calcium concentration for three different values of Diffusion coefficient $D_{Ca}$ = 250 – 350 µm²/s. In both cases as the value of diffusion coefficient increases more numbers of calcium ions get free, hence the calcium concentration increases. Calcium concentration decrease rapidly and finally approaches to $0.1 \mu M$ as we move away from the source.

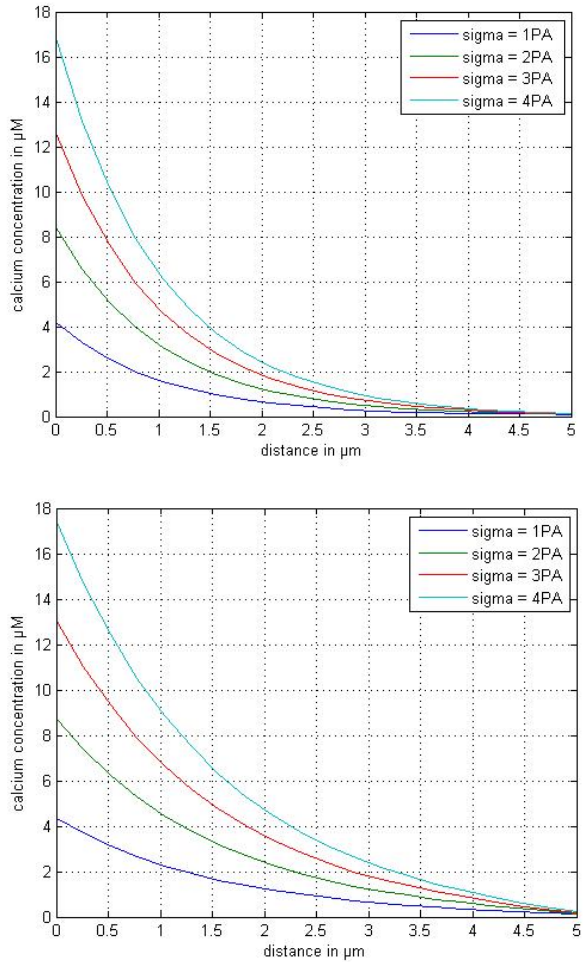

Figure 5 (a) and 5 (b) spatial variation of calcium concentration for different value of influx at boundary

Figure 5 (a) and 5 (b) shows the spatial variation of calcium concentration for four different values of influx. The four different values of influx are $\sigma_{Ca}, 2\sigma_{Ca}, 3\sigma_{Ca}$ and $4\sigma_{Ca}$. In both cases as the value of influx increases more numbers of calcium ions get free, hence the calcium concentration increases. Calcium concentration decrease rapidly up to 3µm and finally approaches to $0.1 \mu M$ as we move away from the source.

## 4. Conclusion

It is observed that ER flux has significant effect calcium concentration gives better central regions little away from the source. The Finite Volume Model is developed here gives us quite interesting results as such models can be developed to generate information about relationship among physical and physiological parameter in the problem and give us better insights and understanding of the chemical signaling phenomena in Astrocytes.

**Brajesh Kumar Jha** is a research scholar in department of Applied Mathematics and Humanities from Sardar Vallabha Bhai National Institute of Technology, Surat. He has completed his M.Sc. in Applied Mathematics from SOMAAS, Jiwaji University, Gwalior. He has number of publication in international journals. His current research interests are Mathematical Modelling and computational biology.

**Dr. Neeru Adlakha** is working as Associat Professor in Department of Applied Mathematics and Humanities from Sardar Vallabha Bhai National Institute of Technology, Surat. She has done Ph.D in mathematics in area of computational biology from jiwaji university, Gwalior. She has won number of awards for her research work. She is ceretary of special group of intrest of Bioinformatics (CSI). She is also member of work group on Bioinformatics and TC-5 of IFIP.

**Dr. M. N. Mehta** is Professor in Mathematics, Deptt. Of Applied Mathematics and Humanities, SVNIT, Surat, Gujarat (INDIA). Former Head of Applied Science and Humanity Deptt. SVNIT, Surat. He did B.Sc., M.Sc., and Ph. D. from South Gujarat University, Surat. He has published 70 Research papers, organized many National and International Conferences, Seminars and workshops. Teaching UG/PG courses since 36 years and wrote some books for UG students. He supervised 14 Ph. D. and some M.Phill candidate. His research interest is Fluid flow through Porous Media.